# Gravitational-Wave Stochastic Background Detection with Resonant-Mass Detectors.

S. Vitale

Department of Physics, University of Trento and Istituto Nazionale di Fisica Nucleare, Gruppo Collegato di Trento, I-38050, Povo, Trento, Italy.

M. Cerdonio

Department of Physics, University of Padova and Istituto Nazionale di Fisica Nucleare, Sez. di Padova, Via Marzolo 8, I-35100 Padova, Italy.

E. Coccia

Department of Physics, University of Roma"Tor Vergata" and Istituto Nazionale di Fisica Nucleare, Sez. di Roma 2 , I-00133 Roma, Italy.

A. Ortolan

Istituto Nazionale di Fisica Nucleare, Laboratori Nazionali di Legnaro, Via Romea 4, I-35020 Legnaro, Padova, Italy.

## Abstract

In this paper we discuss how the standard optimal Wiener filter theory can be applied, within a linear approximation, to the detection of an isotropic stochastic gravitational-wave background with two or more detectors. We apply then the method to the AURIGA-NAUTILUS pair of ultra low temperature bar detectors, near to operate in coincidence in Italy, obtaining an estimate for the sensitivity to the background spectral density of $\approx 10^{-49}$ Hz$^{-1}$, that converts to an energy density per unit logarithmic frequency of $\approx 8 \times 10^{-5} \times \rho_c$, with $\rho_c \approx 1.9 \times 10^{-26}$ kg/m$^3$ the closure density of the Universe. We also show that by adding the VIRGO interferometric detector under construction in Italy to the array, and by properly re-orienting the detectors, one can reach a sensitivity of $\approx 6 \times 10^{-5} \times \rho_c$. We then calculate that the pair formed by VIRGO and one large mass spherical detector properly located in one of the nearby available sites in Italy can reach a sensitivity of $\approx 2 \times 10^{-5} \times \rho_c$ while a pair of such spherical detectors at the same sites of AURIGA and NAUTILUS can achieve sensitivities of $\approx 2 \times 10^{-6} \times \rho_c$.

# 1. Introduction

Resonant-mass gravitational-wave (GW) detectors are potentially interesting for the observation of an isotropic stochastic background of GW, as recently suggested by estimations[1] based on cosmological string models. These models predict for the dilatonic component of that background an almost frequency-independent GW spectral density $S_h(\omega)$ that would put all the detectors under development on a comparable level of sensitivity.

An experiment aimed at detecting a stochastic background of GW implies a correlation technique between the outputs of two or more detectors, possibly distant to each other to avoid spurious correlation of the detector noises. A sub optimal method has been discussed by Michelson[2] with reference to a generic detector, and a quasi-optimal method based on maximum likelihood estimation has been discussed[3] and applied to interferometric detectors. Also a linearized quasi-optimal method has been more recently discussed again in connection with an interferometer pair[4].

Here we first discuss the linearized quasi-optimal estimation method that can be obtained in a straightforward manner from the standard Wiener theory for optimal filtering. We show that it gives the same results as those of ref. 3 for an array of detectors and, when applied to just a pair, of ref. 4, but in a more straightforward way. We then use the method to discuss the potential sensitivity of the pair of ultra low temperature resonant detectors AURIGA-NAUTILUS, which have been tested at two sites[5] at R ≈ 400 Km apart in Italy.

We then extend the calculation to the AURIGA-NAUTILUS-VIRGO[6] array that will operate in Italy.

We finally discuss a possible experimental detection strategy for spherical detectors and estimate the potential sensitivity of the pair formed by VIRGO and one spherical detector and of an array of two of such detectors.

# 2. Isotropic GW stochastic background

Isotropic stochastic background of GW has been widely discussed by many authors[7]. Here, we just summarise some of the results reported and give some useful details omitted in the literature. We assume that the metric tensor perturbation can be written as

$$h_{ij}(\boldsymbol{r},t) = \frac{1}{(2\pi)^3} \iiint h_{ij}(\boldsymbol{k},t) e^{i\boldsymbol{k}\cdot\boldsymbol{r}} \, d^3k \qquad [1]$$

Under the assumption that the background is isotropic and stationary, the two-point two-time auto correlation of the metric tensor $\langle h_{ij}(r,t)h_{kl}(r',t')\rangle$, where the brackets $\langle\ \rangle$ indicate an ensemble average, can be written as (see appendix A)

$$\langle h_{ij}(r,t)h_{kl}(r',t')\rangle = \frac{1}{2\pi} \cdot \int_{-\infty}^{\infty} d\omega S(\omega) e^{-i\omega(t-t')} \times$$
$$\left[\alpha_0(\omega R/c) T^0_{ijkl} + \alpha_1(\omega R/c) T^1_{ijkl} + \alpha_2(\omega R/c) T^2_{ijkl}\right] \quad [2]$$

The functions $\alpha_0(x)$, $\alpha_1(x)$ and $\alpha_2(x)$ in the above equation are given by

$$\alpha_0(x) \equiv \frac{15}{8}\left[\frac{\sin x}{x}\left(1-\frac{5}{x^2}+\frac{9}{x^4}\right)+\frac{\cos x}{x^2}\left(2-\frac{9}{x^2}\right)\right]$$
$$\alpha_1(x) \equiv \frac{15}{8}\left[\frac{\sin x}{x}\left(-1+\frac{1}{x^2}+\frac{3}{x^4}\right)-\frac{\cos x}{x^2}\left(2+\frac{3}{x^2}\right)\right] \quad [3]$$
$$\alpha_2(x) \equiv \frac{15}{8}\left[\frac{\sin x}{x}\left(\frac{6}{x^2}-\frac{12}{x^4}\right)-\frac{\cos x}{x^2}\left(2-\frac{12}{x^2}\right)\right]$$

while the components $T^n_{ijkl}$ of the matrices $\boldsymbol{T^n}$ are given again in appendix A.

Note that, since $T^0_{1111}=1$, and $T^1_{1111}=T^2_{1111}=0$, and since $\lim_{x\to o}\alpha_o(x)=1$, eqs. 2 and 3 give

$$\langle h_{11}(r,t)h_{11}(r,t')\rangle = \frac{1}{2\pi} \cdot \int_{-\infty}^{\infty} d\omega S(\omega) e^{-i\omega(t-t')} \quad , \quad [4]$$

which is a standard normalisation in the context of signal analysis.

The antenna pattern of any GW detector, in the approximation of detector size that is small compared with the gravitational wavelength, can be described[8] by a symmetric traceless tensor $\boldsymbol{D}$. By this we mean that the effective metric perturbation sensed by the detector is $h(t)=\sum_{ij=1}^{3}D_{ij}h_{ji}(t)$ with $h_{ij}(t)$ the incoming wave. The explicit expression for the components $D_{ij}$ of $\boldsymbol{D}$, for the lowest longitudinal mode of a cylindrical GW antenna, is $D_{ij} = n_i n_j - \delta_{ij}/3$, where $\boldsymbol{n}$, with components $n_i$, is the unit vector parallel to its axis. The lowest five degenerate quadrupole modes of a spherical detector have, instead, the following coupling tensors[9] $\boldsymbol{D^m}$ (m=-2,...,+2) : $\boldsymbol{D^0} = \frac{\sqrt{3}}{6}\left(2e^+_{xz} - e^+_{xy}\right)$, $\boldsymbol{D^1} = -\frac{1}{2}e^\times_{yz}$, $\boldsymbol{D^{-1}} = -\frac{1}{2}e^\times_{xz}$, $\boldsymbol{D^2} = \frac{1}{2}e^+_{xy}$ and $\boldsymbol{D^{-2}} = -\frac{1}{2}e^\times_{xy}$, where the tensors $e^{\times,+}_{\alpha\beta}$ are defined in appendix A.

Finally, for an interferometric antenna with its arms in the direction of the unit vectors $\boldsymbol{n}$ and $\boldsymbol{m}$, we have $D_{ij} = (n_i n_j - m_i m_j)/2$.

The cross correlation $R^{ab}_h(t-t') \equiv \langle h^a(t) h^b(t') \rangle$ between the effective metric perturbation signals $h^a(t)$ and $h^b(t)$ sensed by two detectors a and b located at $\boldsymbol{r}$ and $\boldsymbol{r'}$, due to the stochastic GW background, can be written as

$$R^{ab}_h(t-t') = \frac{1}{2\pi} \cdot \int_{-\infty}^{\infty} d\omega\, S(\omega) e^{-i\omega(t-t')} \times \left[\Omega_0\, \alpha_0(\omega R/c) + \right. \quad [5]$$
$$\left. + \Omega_1\, \alpha_1(\omega R/c) + \Omega_2\, \alpha_2(\omega R/c)\right]$$

where $\Omega_\alpha \equiv \sum_{ijkl=1}^{3} D^a_{ij} D^b_{lk} T^\alpha_{ijkl}$ are three constants that only depend on the nature of the detectors, their locations, and their relative orientations.

Let us now take two resonant bars. The force acting on antenna a placed in $\boldsymbol{r}_a$ is

$$F_a(t) = \frac{1}{2} m_a l_a \frac{d^2 h^a(t)}{dt^2} \quad [6]$$

where $m_a$ is the effective mass (1/2 of the physical mass M) and $l_a$ is the effective length (for a cylinder $l_a = (4/\pi^2)L$, with L the physical length of the cylinder). For two parallel oriented bars (the orientation that always maximises the correlation) we get

$$\langle F_a(t) F_b(t') \rangle =$$

$$= m_a m_b l_a l_b \frac{1}{2\pi} \int_{-\infty}^{\infty} S(\omega) \omega^4 e^{-i\omega(t-t')} \Theta(\vartheta, \omega R/c) d\omega \qquad [7]$$

where:

$$\Theta(\vartheta, x) = \frac{\alpha_0(x)}{8}[7 + 4\cos(2\vartheta) + 5\cos(4\vartheta)] + $$
$$+ \frac{\alpha_1(x)}{2}[1 + 2\cos(2\vartheta) + \cos(4\vartheta)] + \frac{\alpha_2(x)}{2}[1 - \cos(4\vartheta)] \qquad [8]$$

here, $\vartheta$ is the angle between the direction of the detector axis (the same direction for both) and the straight line joining the sites where the detectors are located. The behaviour of the function $\Theta(\vartheta, x)$ is reported in figure 1 with $\vartheta$ as a parameter.

It is worth pointing out that, if the detectors have both to be horizontal and still be exactly parallel, $\vartheta$ can only be $\vartheta = 90°$. However, as in the following sections we will focus on observatories where $R \ll R_\oplus$, with $R_\oplus$ the radius of the Earth, it is also worth noticing that in that case, whatever the value of $\vartheta$, the detectors can be made approximately parallel within an angle $\delta \approx R/R_\oplus$.

For two spheres, each of the five degenerate quadrupole mode of the sphere acts as an independent detector[9]. We can write the GW force acting on the m-th mode as

$$F_a^m(t) = \frac{1}{2} m_a l_a \frac{d^2 h_m^a(t)}{dt^2}, \qquad [9]$$

where now the effective mass and length refer to the mode ($m_a$ is the physical mass of the sphere and $l_a \approx 0.3$ of the diameter of the sphere[10]) and $h_m^a(t)$ is obtained by contraction with the proper tensor $\boldsymbol{D^m}$. The cross correlation among the forces acting on the five modes of the two spheres separated by a distance R, assuming that all the modes are defined with respect to the same coordinate system, is then

$$\langle F_a^m(t) F_b^{m'}(t') \rangle =$$

$$= \left[ m_a l_a m_b l_b \frac{1}{2\pi} \int_{-\infty}^{\infty} e^{-i\omega(t-t')} S(\omega) \omega^4 \, \Xi(m, \omega R/c) \, d\omega \right] \delta^{mm'} \quad [10]$$

where

$$\Xi(m,x) = \begin{cases} \frac{3}{2}[\alpha_0(x) + \alpha_1(x)] & m = 0 \\ \alpha_2(x) & m = \pm 1 \\ \frac{1}{2}[\alpha_0(x) - \alpha_1(x)] & m = \pm 2 \end{cases} \quad [11]$$

Note that for any value of m, $\Xi(m,x)$ has the same limit of 3/4 for $R \to 0$ and thus the sum over the five modes of two co-located spheres gives the usual 15/4 factor coming from the solid-angle average of the energy emitted by a GW source. Figure 2 reports the behaviour of $\Xi(m,x)$ as a function of the dimensionless quantity $x = \omega R/c$. The correlation of the $m = 0, \pm 1$ modes decays faster than the $m = \pm 2$; therefore in a correlation experiment with two non co-located spheres ($\omega R/c \geq 5$) we have only these two modes relevant for the detection of the GW stochastic background.

## 3. The linearized estimation method

The stochastic force resulting from the GW background adds just an extra contribution to the Gaussian noise of a single detector. As a consequence, a measurement with a single detector has to rely on an a priori estimate of the detector noise that has to be subtracted from the data, a procedure whose accuracy is always highly questionable.

A much safer method is obviously to cross correlate the outputs of two or more detectors with no common source of noise. Correlation of noise sources, such as electromagnetic interference, seismic noise, etc., is expected to decay with the distance between the detectors. Cross correlation between detectors located far apart is then the most advisable procedure to estimate the background.

The Wiener theory of optimal estimation can be applied both to deterministic signals and to stochastic processes buried in detector noise. In this section we show that through a suitable small-signal linearization, a Wiener-like method can be applied to a cross-correlation estimate of the amplitude of a stochastic background signal of known power spectrum driving a set of noisy detectors.

Assume then that the power spectrum of the GW background $S(\omega)$ in eq. 2 can be written as $S(\omega) = A^2 \tilde{S}(\omega)$, where $\tilde{S}(\omega)$ is a known function of $\omega$, while the amplitude A is unknown and has to be estimated. $\tilde{S}(\omega)$ can be a flat function of $\omega$, as suggested by some recent string theory[1], or the power law $\tilde{S}(\omega) \propto \omega^{-\gamma}$, which is predicted by the standard inflationary cosmology[7]. The output data streams of a set of N detectors can then be written:

$$x^a(t) = \eta^a(t) + As^a(t) \qquad 1 \leq a \leq N, \qquad [12]$$

where $\eta^a(t)$ are stochastic processes that describe the noise in each detector and $s^a(t)$ is the signal due to the stochastic GW background.

The statistical properties of the Gaussian processes in eq. 2 can be summarised as follows:

$$\begin{cases} \langle \eta^a(t) \rangle = \langle s^a(t) \rangle = 0 \\ \langle \eta^a(t) \eta^b(t') \rangle = R_n^a(t-t')\delta_{ab} & 1 \leq a,b, \leq N \\ \langle s^a(t) s^b(t') \rangle = R_s^{ab}(t-t') \\ \langle n^a(t) s^b(t') \rangle = 0 \end{cases} \qquad [13]$$

The correlation $R_s^{ab}(t-t')$ among the GW signals available at the output of the detectors are

$$R_s^{ab}(t-t') = \int_0^\infty dt'' H^a(t'') \int_0^\infty dt''' H^b(t''') \tilde{R}_h^{ab}(t-t'-t''+t''') \qquad [14]$$

where $\tilde{R}_h^{ab}(t-t')$ is related to $\tilde{S}(\omega)$ by the same eq. 5 that relates $R_h^{ab}(t-t')$ to $S(\omega)$. The response functions $H^a(t)$ of the $a^{th}$ detector translate the input signal $h^a(t)$ into the output signal

$$As^a(t) = \int_0^\infty H^a(t') h^a(t-t') dt'. \qquad [15]$$

We now look for an optimal estimator of the spectrum amplitude by writing down the most general bilinear combination of the outputs of N detectors:

$$\hat{A}^2 = \sum_{a,b=1}^{N} \int_{-T}^{T} dt \int_{-T}^{T} dt'\, g_{ab}(t,t') x^a(t) x^b(t') , \qquad [16]$$

where the "filter" functions $g_{ab}(t,t')$ have to be chosen such that $\hat{A}^2$ is an unbiased estimator:

$$\langle \hat{A}^2 \rangle = A^2 \qquad [17]$$

and that its variance $\sigma_{\hat{A}^2}^2 \equiv \langle (\hat{A}^2)^2 \rangle - \langle \hat{A}^2 \rangle^2$ is minimal.

These requests lead in general to a non-linear problem. However, we show in appendix B that if

$$A s^a(t) << n^a(t), \qquad [18]$$

which is likely to be the case for actual detectors, then the problem can be linearized, and $g_{ab}(t,t')$ obeys the integral equation

$$2 \int_{-T}^{T} dt'' \int_{-T}^{T} dt'''\, g_{ab}(t'',t''') R_n^a(t-t'') R_n^b(t'-t''') = -\frac{\lambda}{2} R_s^{ab}(t-t') \qquad [19a]$$

for $a \neq b$; while

$$g_{ab}(t-t') = 0 \qquad \text{for } a = b \qquad [19b]$$

The Lagrange multiplier $\lambda/2$ in eq. 19a is obtained from

$$\sum_{a,b=1}^{N} \int_{-T}^{T} dt \int_{-T}^{T} dt'\, g_{ab}(t,t') R_s^{ab}(t'-t) = 1 \qquad [20]$$

and we can also show that:

$$\sigma^2_{\hat{A}^2} \equiv -\frac{\lambda}{2} \qquad [21]$$

Eq. 19a can be solved numerically. However much of the information about the solution can be obtained by assuming that $g_{ab}(t,t')$ decays rapidly as a function of $|t-t'|$. To be more specific, if the optimal bandwidths of the detectors are of order $1/\tau$ and if the light travel time is $\approx R/c$, it is then reasonable to assume that $g_{ab}(t,t') \to 0$ if $\{|t-t'|-R/c\}/\tau >> 1$ and then choose to integrate on a time $2T>>2\tau+R/c$. If this is the case, we can take $T \approx \infty$ in eq 19a and solve it in terms of Fourier transforms:

$$g_{ab}(\omega) \approx -\frac{\lambda}{2}\frac{S_s^{ab}(\omega)}{2S_n^a(\omega)S_n^b(\omega)}, \qquad [22]$$

where

$$g_{ab}(t,t') \approx g_{ab}(t-t') = \frac{1}{2\pi}\int_{-\infty}^{\infty} g_{ab}(\omega) e^{i\omega(t-t')} d\omega, \qquad [23]$$

and where we have defined $S_n^a(\omega)$ and $S_s^{ab}(\omega)$ as the Fourier transforms of $R_n^a(t-t')$ and $R_s^{ab}(t-t')$, respectively.

To evaluate the variance of the estimate, we can then substitute this solution in eq. 20:

$$\sigma^{-2}_{\hat{A}^2} =$$

$$= \sum_{a \neq b}\frac{1}{(2\pi)^2}\int_{-\infty}^{+\infty}\int_{-\infty}^{+\infty} d\omega d\omega' \frac{S_s^{ab}(\omega)S_s^{ab}(\omega')}{2S_n^a(\omega)S_n^b(\omega)}\int_{-T}^{T}\int_{-T}^{T} dt dt' \, e^{i[(t-t')\cdot(\omega-\omega')]} = \qquad [24]$$

$$= \sum_{a \neq b}\frac{T}{2\pi}\int_{-\infty}^{+\infty}\int_{-\infty}^{+\infty} d\omega d\omega' \frac{S_s^{ab}(\omega)S_s^{ab}(\omega')}{S_n^a(\omega)S_n^b(\omega)} \frac{\sin[(\omega-\omega')T]}{(\omega-\omega')T}\delta_T(\omega-\omega')$$

where $\delta_T(\omega) \equiv \frac{1}{2\pi}\int_{-T}^{T} e^{i\omega t} dt = \frac{1}{\pi}\frac{\sin(T\omega)}{\omega}$ is a finite-time approximation to the Dirac $\delta$ function, which reduces to $\delta(\omega)$ in the limit $T \to \infty$. If we assume again that the observation time $T$ is large enough to have null correlation functions for $|t-t'|>T$, then $\delta_T(\omega)$ is a sharply peaked function in a very small frequency range compared to the scales on which the functions $S_s^{ab}(\omega)$ and $S_n^a(\omega)$ change. In this case then eq. 24 gives

$$\sigma_{\hat{A}^2} \cong \left[ \sum_{a<b} \frac{T}{\pi} \int_{-\infty}^{+\infty} d\omega \frac{\left[S_s^{ab}(\omega)\right]^2}{S_n^a(\omega)S_n^b(\omega)} \right]^{-1/2} . \qquad [25]$$

In eq 25 we have also changed from the sum over $a \neq b$ to that over $a<b$, which contains half the terms. Note then that, if we define

$$\sigma_{ab} = \left[ \frac{T}{\pi} \int_{-\infty}^{+\infty} d\omega \frac{\left[S_s^{ab}(\omega)\right]^2}{S_n^a(\omega)S_n^b(\omega)} \right]^{-1/2} , \qquad [26]$$

which is the uncertainty of the estimate that only uses the data from detectors a and b, eq. 25 becomes

$$\frac{1}{\sigma_{\hat{A}^2}^2} = \sum_{a<b} \frac{1}{\sigma_{ab}^2} \qquad [27]$$

The result in eq. 25 can be further clarified: let us call $H^a(\omega)$ the transfer function of the $a^{th}$ detector, i.e. the Fourier transform of the response function $H^a(t)$ appearing in eq. 15. Then clearly $S_s^{ab}(\omega) = \tilde{S}_h^{ab}(\omega) |H^a(\omega)|^2 |H^b(\omega)|^2$, where $\tilde{S}_h^{ab}(\omega)$ is the Fourier transform of $\tilde{R}_h^{ab}(t-t')$ in eq 15. By defining the equivalent input detector noise as $S_a^h(\omega) = S_a^n(\omega) / |H^a(\omega)|^2$, eq. 25 can be rewritten as:

$$\sigma_{\hat{A}^2}^2 = \left[ \sum_{a<b} \frac{T}{\pi} \int_{-\infty}^{\infty} d\omega \frac{\left[\tilde{S}_h^{ab}(\omega)\right]^2}{S_h^a(\omega)S_h^b(\omega)} \right]^{-1} , \qquad [28]$$

or, by using eq. 2 and the definition of $\tilde{S}(\omega)$

$$\sigma^2_{\hat{A}^2} = \left[ \sum_{a<b} \frac{T}{\pi} \int_{-\infty}^{\infty} d\omega \times \right.$$

$$\left. \times \frac{\{\tilde{S}(\omega)[\Omega_0 \alpha_0(\omega R/c) + \Omega_1 \alpha_1(\omega R/c) + \Omega_2 \alpha_2(\omega R/c)]\}^2}{S_h^a(\omega) S_h^b(\omega)} \right]^{-1} \quad [29]$$

Note that eq. 27 agrees with the results of ref. 3, which have been derived with a different approach.

The weight function $g_{ab}(t-t')$, which is just the optimal bilinear filter for the search of the stochastic background, can be explicitly evaluated by solving eq. 19a in the Fourier space:

$$g_{ab}(\omega) = \sigma^2_{\hat{A}^2} \times$$

$$\times \frac{\tilde{S}(\omega)[\Omega_0 \alpha_0(\omega R/c) + \Omega_1 \alpha_1(\omega R/c) + \Omega_2 \alpha_2(\omega R/c)]}{S_h^a(\omega) S_h^b(\omega)} \quad [30]$$

The filter depends on the shape of power spectrum $\tilde{S}(\omega)$, but the dependence should be very weak for any detector array that includes resonant detectors. In fact such detectors have comparatively narrow bands, so if $S_h^a(\omega)$ is its equivalent noise input spectrum, then $S_h^a(\omega)$ grows very rapidly outside the post detection band centred at some centre frequency $\omega_0$ in the kHz range. As a consequence, the template in eq. 30 decreases rapidly outside the same band. In this limit, $\tilde{S}(\omega)$ can be approximated by a constant $\tilde{S}(\omega) \approx \tilde{S}(\omega_0)$.

If, instead, the array includes only wide bandwidth detectors, a set of filters with $\tilde{S}(\omega) \propto \omega^{-\gamma}$ should be constructed and the data processed with the different choices of $\gamma$ predicted by different cosmological models. The filters that maximise the signal-to-noise ratio $A^2/\sigma_{\hat{A}^2}$ gives then an estimate of $\gamma$.

## 4. Implications for AURIGA, NAUTILUS, VIRGO and large mass spherical detectors.

In this section, we use eq. 29 to discuss the sensitivity of the AURIGA and NAUTILUS pair (soon to be operated in coordinate coincidence) to the GW. stochastic background. We also discuss the expected performance of the AURIGA-NAUTILUS-VIRGO array and the sensitivity of arrays including one or more large mass spheres.

Let us consider the simplest model for a resonant antenna, in close analogy with the viewpoint suggested by Giffard[11]. The $a^{th}$ antenna is considered as a simple harmonic oscillator of mass $m_a$ and length $l_a$ with resonant angular frequency $\omega_a$ and decay time $\tau_a$, excited by the force. The position $x^a(t)$ of the oscillator mass is read by a suitable position transducer. The transfer function from the input force to the output displacement is

$$H_a(\omega) = \frac{l_a}{2m_a} \frac{\omega^2}{\omega_a^2 - \omega^2 + i\omega/\tau_a} \qquad [31]$$

Within this model the oscillator is driven into random motion by the sum of the brownian noise force and the back-action noise force of the position transducer. The position transducer also contributes with an additive position noise $x_n^a(t)$. Both the total noise force $f_n^a(t)$ and the additive displacement noise $x_n^a(t)$ are assumed to have white spectra with values $S_f^a$ and $S_x^a$ respectively. $S_f^a$ and $S_x^a$ can be parametrized as:

$$S_f^a \equiv N^a h k_n^a$$
$$S_x^a \equiv \frac{N^a h}{k_n^a} \qquad [32]$$

where $N^a$ represents the noise quantum number, which has to be $N^a \geq 1/2$, and $k_n^a = \sqrt{S_f^a/S_x^a}$ is a noise "stiffness". These quantities have an immediate physical meaning because they directly relate to the antenna burst sensitivity $h_{min}^a$ and post detection bandwidth $\Delta\omega_{pd}^a$ respectively. Here, $h_{min}^a$ is defined as the amplitude of the burst, of centre frequency $\approx \omega_a$ and duration $\approx 1/\omega_a$, which gives a signal to noise ratio of one. The relation of $N^a$ and $k_n^a$ to $h_{min}^a$ and $\Delta\omega_{pd}^a$ is:

$$h_{min}^a \approx \frac{1}{l_a}\left(\frac{2N^a \hbar}{m_a \omega_a}\right)^{1/2} \quad ; \quad \Delta\omega_{pd}^a \approx \frac{k_n^a}{m_a \omega_a} \qquad [33]$$

The total noise at the antenna output is given by the sum of a narrow band and a wide band contributions:

$$S_n^a(\omega) = S_x^a + \frac{S_f^a}{m_a^2} \frac{1}{\left(\omega_a^2 - \omega^2\right)^2 + \left(\omega/\tau_a\right)^2} \qquad [34]$$

This noise can be referred to the antenna input, as if it were a spectrum, dividing $S_n^a(\omega)$ by the squared antenna transfer function. The resulting noise spectrum is shaped as the inverse of a lorentzian function and can be written as

$$S_h^a(\omega) = \frac{\left(2h_{min}^a\right)^2}{\Delta\omega_{pd}} \frac{\left[\left(\hat{\omega}_a^2 - \omega^2\right)^2 + \left(\omega\Delta\omega_{pd}^a\right)^2\right]}{\omega^4} \qquad [35]$$

where $\hat{\omega}_a^4 = \omega_a^4\left(1 + \frac{\left(k_n^a\right)^2}{4m_a^2\omega_a^4}\right) \approx \omega_a^4$ is the frequency at which the noise reaches a minimum and which coincides in practice with the resonance frequency of the antenna.

$S_h^a(\omega)$ reaches its minimum value at $\omega \approx \hat{\omega}_a$:

$$\left[S_h^a(\omega)\right]_{min} \approx S_h^a(\omega_a) \approx \frac{\left(2h_{min}^a\right)^2}{\omega_a^2} \Delta\omega_{pd}^a \approx \frac{8k_B T^*}{m_a l_a^2 \omega_a^3 Q_a} \qquad [36]$$

where $Q_a = \omega_a \tau_a$ and $T^*$ is the temperature of the antenna.

## 4.1 Two bars

Let us consider the case of two bars. Using the physical length $L_a$ and mass $M_a$ eq. 36 gives for a single bar:

$$S_h^a(\omega_a) \approx 1.3 \times 10^{-45} \times$$

$$\times \left(\frac{L_a}{3\,m}\right)^{-2} \left(\frac{M_a}{2300\,Kg}\right)^{-1} \left(\frac{T^*}{50\,mK}\right) \left(\frac{1\,KHz}{\nu_a}\right)^3 \left(\frac{Q_a}{10^7}\right)^{-1} Hz^{-1} \qquad [37]$$

To evaluate the sensitivity of a pair of bars to the stochastic background we substitute eq. 35 in eq. 28, obtaining:

$$\sigma_{\hat{A}^2}^2 \approx \left[ \frac{T}{\pi} \int_{-\infty}^{\infty} \frac{\omega_a^2 \omega_b^2}{\Delta\omega_{pd}^a \Delta\omega_{pd}^b \left(4 h_{min}^a h_{min}^b\right)^2} \times \right.$$

$$\left. \times \frac{\omega^8 \tilde{S}^2(\omega) \Theta^2(\vartheta, \omega R/c) d\omega}{\left[\left(\omega_a^2 - \omega^2\right)^2 + \left(\omega \Delta\omega_{pd}^a\right)^2\right]\left[\left(\omega_b^2 - \omega^2\right)^2 + \left(\omega \Delta\omega_{pd}^b\right)^2\right]} \right]^{-1} \qquad [38]$$

To carry out our calculations, we now assume that the detectors are parallel and that they have the same sensitivity, the same post detection bandwidth but, to take into account the real physical situation, we allow for slightly different resonant frequencies. For $\left(\omega_a/\Delta\omega_{pd}^a\right)^2 \gg 1$, the integral in eq. 38 can be approximated by

$$\sigma_{\hat{A}^2} \approx \left[ \frac{\left[S_h^a\right]_{min} \left[S_h^b\right]_{min}}{2T\Delta\omega_{pd}^a} \frac{1 + \left(|\omega_a - \omega_b|/\Delta\omega_{pd}^a\right)^2}{\Theta^2(\vartheta, \overline{\omega} R/c)} \right]^{1/2}, \qquad [39]$$

where $\overline{\omega} \equiv (\omega_a + \omega_b)/2$ is the mean frequency of the two antennae and where we have taken $\tilde{S}(\omega = \omega_a) = 1$.

For $R = 400$ Km, which is the distance of the AURIGA-NAUTILUS pair, and with an average frequency of $\overline{\omega} \cong (2\pi \cdot 920)$ rad/s, $\overline{\omega} R/c \cong 7.7$ while the expected sensitivity is $\left[S_h^a\right]_{min} \approx 1.7 \times 10^{-45}$ Hz$^{-1}$. Taking $\vartheta = 90°$ one gets:

$$\sigma_{\hat{A}^2} \approx 10^{-49} \left(\frac{2T}{1\,\text{year}}\right)^{-1/2} \left(\frac{\omega_a/\Delta\omega_{pd}^a}{30}\right)^{-1/2} \text{Hz}^{-1}, \qquad [40]$$

provided that the detuning of the resonance frequencies is small compared with the optimal bandwidth. This last condition is, for instance, matched with 25 % approximation if $|\omega_a - \omega_b|/\Delta\omega_{pd}^a \leq 1/2$. With an effective bandpass > 20 Hz, this implies that the two detectors have to be matched within 10 Hz, which appears to be feasible.

However, in order to have both detectors parallel to each other and as parallel as possible to the other cryogenic detectors already in operation[12], AURIGA and NAUTILUS are presently oriented with $\vartheta \cong 52°$. This value gives a sensitivity of roughly a factor of 2 worse than that in eq. 40 (Fig.1).

To reorient the detectors is technically feasible, but it is doubtful whether the factor of 2 is worth the loss of parallelism with the remaining detectors.

*4.2 Two bars and one interferometer.*

We now discuss the potential sensitivity of the AURIGA-NAUTILUS-VIRGO array, assuming for VIRGO the planned position (Lat = 10° 30' E, Long = 43° 40' N) and orientation (one arm at 26°, the other at 296°). There are many noise sources in an interferometric antenna that have been carefully estimated[13]. However, the resulting total noise power spectrum in the frequency range of interest here, i.e., around 1 kHz, is dominated by the shot-noise contribution:

$$S_h^{virgo}(\omega) \approx S_{ho}^{virgo}\left(\frac{\omega}{\omega_0}\right)^2 \qquad 400\text{ Hz} < \omega/2\pi < 4000\text{ Hz}, \qquad [41]$$

where, from the published curves of the expected noise spectrum[13], one can estimate $S_{ho}^{virgo} \cong 1.6 \times 10^{-45}$ Hz$^{-1}$ if $\omega_0$ is taken $\omega_0 = 2\pi 920$ rad/s. This gives a noise value close to the minimum value of the input noise of AURIGA or NAUTILUS.

One can apply eq. 5 to the pairs AURIGA-VIRGO and NAUTILUS-VIRGO, which have intersite distances of $\approx 220$ Km and $\approx 260$ Km respectively, corresponding to $\omega_a R/c \cong 4.2$ and $\omega_a R/c \cong 5$ at a frequency of 920 Hz. The correlation value corresponding to the AURIGA-VIRGO pair turns to be ≈2 greater than that for NAUTILUS-VIRGO.

The sensitivity to the stochastic background for each bar-interferometer pair and for one year of integration is obtained by substituting eqs. 40 and 35 in eq. 28 and by integrating:

$$\sigma_{\hat{A}^2} \cong \left( \frac{\left[S_h^a\right]_{min} S_{ho}^{virgo}}{T\Delta\omega_{pd}^a} \frac{1}{\Theta^2(\omega R/c)} \right)^{1/2}, \qquad [42]$$

where we can see that the overall bandwidth is set by that of the resonant antenna. The resulting sensitivity for the pair AURIGA-VIRGO is $\sigma_{\hat{A}^2} \cong 2 \times 10^{-49} \, \text{Hz}^{-1}$, assuming for AURIGA $\omega_a/\Delta\omega_{pd} = 30$ and the present orientation; while for the NAUTILUS VIRGO pair, again with the present orientation, $\sigma_{\hat{A}^2} \cong 3.5 \times 10^{-49} \, \text{Hz}^{-1}$. For the array as a whole, using eq. 27, these values give $\sigma_{\hat{A}^2} \cong 1.3 \times 10^{-49} \, \text{Hz}^{-1}$. With the bar detectors oriented with $\vartheta = 90°$, a choice that maximise the overall sensitivity, one gets instead $\sigma_{\hat{A}^2} \cong 8 \times 10^{-50} \, \text{Hz}^{-1}$.

Although the improvement in sensitivity with respect to the pair of bars is almost negligible, such a three-detectors detection would be of paramount importance in ruling out spurious effects.
.

### *4.3 Two spheres*

Finally we study the performances of two spherical detectors, which have been recently proposed as possible next-generation resonant antennas.

Eq. 38 only involves the detectors post detection bandwidth and input noise. We can easily compare the sensitivity of a bar to that of a sphere made of the same material, and with the same resonant frequency, as[14]:

$$\left[S_h^{bar}\right]_{min} \approx 1.17 \left[S_h^{sphere}\right]_{min} (M_s/M_b), \qquad [43]$$

where $M_s$ and $M_b$ are the physical masses of the sphere and the bar respectively. For a sphere of diameter D, made of the same material of the present bars (Al 5056) we find

$$S_h^a(\omega_a) \approx 7 \times 10^{-47} \times$$

$$\times \left(\frac{D}{3\,m}\right)^{-2} \left(\frac{M_s}{38000\,Kg}\right)^{-1} \left(\frac{T^*}{50\,mK}\right) \left(\frac{1\,KHz}{\nu_a}\right)^3 \left(\frac{Q_a}{10^7}\right)^{-1} Hz^{-1} \quad [44]$$

Under the same approximation used for eq.38, the sensitivity of a sphere pair, for each mode of the sphere, can be written as:

$$\sigma_{\hat{A}^2}^m \approx \left[\frac{[S_h^a]_{min}[S_h^b]_{min}}{2T\Delta\omega_{pd}^a} \frac{1+(|\omega_a-\omega_b|/\Delta\omega_{pd}^a)^2}{\Xi^2(m,\overline{\omega}R/c)}\right]^{1/2}, \quad [45]$$

If we consider two co-located spheres, the overall sensitivity is higher by a factor $\sqrt{5}$ with respect to that for a single mode, since we can add the signals on the five modes of the sphere. If the spheres are far apart, fig. 2 shows that only for the $m = \pm 2$ modes is the decay of the correlation with distance as slow as that for the bars. For the other modes, the correlation decays much faster, so for x>5, the overall sensitivity is higher only by a factor of $\approx \sqrt{2}$ with respect to that for a single mode.

The expected sensitivity of two co-located 3-m-diameter sphere, made of Al 5056, is

$$\sigma_{\hat{A}^2} \approx 4 \times 10^{-52} \left(\frac{T^*}{1\,year}\right)^{-1/2} \left(\frac{\omega_a/\Delta\omega_{pd}^a}{30}\right)^{-1/2} Hz^{-1} \quad [46]$$

if one takes the weighted average of all the modes. As all the parameters in eq. 45 are the same for the five modes except for the coefficients $\Xi^2(m,\overline{\omega}R/c)$, the error on the weighted average is given by the same eq. 45 by replacing $\Xi^2(m,\overline{\omega}R/c)$ with the sum $\sum_{m=1}^{5} \Xi(m,\overline{\omega}R/c)$ which is reported in Fig. 3. For two spheres at the AURIGA and NAUTILUS sites, $\overline{\omega}R/c \cong 7.7$, one would get then $\sigma_{\hat{A}^2} \approx 2 \times 10^{-51} \left(\frac{2T}{1\,year}\right)^{-1/2} \left(\frac{\omega_a/\Delta\omega_{pd}^a}{30}\right)^{-1/2} Hz^{-1}$.

This figure may improve by changing the sphere material and/or increasing the sphere diameter. For instance if one is able to fabricate two 4-m-diameter copper alloy sphere (250 tons), the above reported sensitivity reaches about 4 10$^{-52}$ Hz$^{-1}$.

*4.4 One sphere and one interferometer*

In fig. 4 we show the correlation function for one interferometer and a sphere. The function is obtained by summing up the contribution coming from all the sphere modes. If the sphere and the interferometer are not at the same site, then the function depends on the angle θ between one of the interferometer arm and the line joining the two sites. For a sphere and an interferometer like VIRGO located at the same site, the figure gives:

$$\sigma_{\hat{A}^2} \approx 8 \times 10^{-51} \left( \frac{2T}{1\,\text{year}} \right)^{-1/2} \left( \frac{\omega_a / \Delta\omega_{pd}^a}{30} \right)^{-1/2} \text{Hz}^{-1} \qquad [47]$$

As the site of VIRGO is fixed, one can try the exercise to locate a sphere in one of the three major laboratories available close to VIRGO. Those are the AURIGA and NAUTILUS sites and the large underground laboratory of Gran Sasso[15]. For these laboratories $\overline{\omega}R/c$ and θ take the values $\overline{\omega}R/c \cong 4.2$, $\theta \cong 4°$, $\overline{\omega}R/c \cong 5$, $\theta \cong 68.7°$ and $\overline{\omega}R/c \cong 5.3$, $\theta \cong 90°$ respectively. With T=1 year and $\omega_a / \Delta\omega_{pd}^a = 30$, these figures give sensitivities that are respectively: $\sigma_{\hat{A}^2} \approx 2.5 \times 10^{-50}\,\text{Hz}^{-1}$ for the AURIGA site, $\sigma_{\hat{A}^2} \approx 5 \times 10^{-50}\,\text{Hz}^{-1}$ for that of NAUTILUS and $\sigma_{\hat{A}^2} \approx 6 \times 10^{-50}\,\text{Hz}^{-1}$ for the Gran Sasso Laboratory.

## 5. Conclusions

The sensitivities reported in this paper can be expressed in term of the ratio $\Omega_{GW}(\omega)$ of the mass-energy density per unit logarithmic frequency of the GW stochastic background to the closure density $\rho_c$ of the universe. In fact the spectrum of the stochastic background S(ω) can be written as[16]:

$$S(\omega) = \frac{16 G \pi^2}{\omega^3} \Omega_{GW}(\omega) \rho_c \cong 8 \times 10^{-46} \Omega_{GW}(\omega) \left( \frac{2 \times \pi \times 1\,\text{kHz}}{\omega} \right)^3$$

An experiment involving at least one resonant detector, bar or sphere, operating at 1 kHz would than measure $\Omega_{GW}(\omega)$ with an uncertainty:

$$\sigma_{\Omega_{GW}} \approx 8 \times 10^{-6} \left( \frac{\sigma_{\hat{A}^2}}{10^{-50}} \right)$$

where $\rho_c = 1.9 \times 10^{-26}$ Kg/m$^3$ has been assumed. The sensitivities discussed in sec. 4 can be then recast in terms of $\Omega_{GV}(\omega)$ according to table I.

| Table I |||
|---|---|---|
| Estimated sensitivities of various detectors arrays. |||
| Detectors array | $\sigma_{\hat{A}^2} \left( Hz^{-1} \right)$ | $\sigma_{\Omega_{GW}}$ |
| AURIGA-NAUTILUS present orientation | 2×10$^{-49}$ | 2×10$^{-4}$ |
| AURIGA-NAUTILUS best orientation | 10$^{-49}$ | 8×10$^{-5}$ |
| AURIGA-NAUTILUS-VIRGO present orientation | 1.3×10$^{-49}$ | 10$^{-4}$ |
| AURIGA-NAUTILUS-VIRGO best orientation | 8×10$^{-50}$ | 6×10$^{-5}$ |
| VIRGO and one 38 ton sphere | 2.5×10$^{-50}$ | 2×10$^{-5}$ |
| Two 38 tons spheres at AURIGA and NAUTILUS sites | 2×10$^{-51}$ | 2×10$^{-6}$ |

These figures have to be compared with experimental limits already existing. Microwave background measurements by COBE[17] limit $\Omega_{GW}$ to about 10$^{-8}$ but just at very low frequencies (<10$^{-15}$ Hz). Pulsar timing observation also give limits < 10$^{-9}$ but again for very low frequencies ( ≈ 10$^{-8}$ Hz)[18]. An estimate from earth quadrupole oscillation[19] in the mHz range gives $\Omega_{GW}$< 3.

Direct experimental measurements on ground in the kHz range have been performed in the past with sensitivities significantly lower than those indicated in table I. A pioneering measurement with split bars[20] gave a limit of $\Omega_{GW}$< 3×10$^3$ in the kHz range. More recent estimates from the background noise of a cryogenic detector[21] give $\Omega_{GW}$< 3×10$^2$ .

Table I then shows that already in the near future experiments involving resonant mass detectors can provide unprecedented upper limits to the GW stochastic background at kHz frequencies. Future sensitivities of bars-interferometer and spheres experiments may hope to go near or beyond the limit put by nucleosinthesys considerations at $\Omega_{GW} \approx 10^{-5}$ and to detect such fundamental cosmic signals.

## Aknowledgments

We thank R. Brustein, M. Gasperini and G. Veneziano for stimulating our interest in this subject and G. Pizzella and E. Picasso for helpful discussions.

## Appendix A

The coefficient of the expansion in eq. 1 can be written as a superposition of in-going and out-going monochromatic plane waves,

$$h_{ij}(\mathbf{k},t) = h_{ij}^{\rightarrow}(\mathbf{k})e^{-ikct} + h_{ij}^{\leftarrow}(\mathbf{k})e^{+ikct}, \quad [A1]$$

where we have denoted by k the modulus of $\mathbf{k}$. Since the metric has to be a real function, we have $h_{ij}(\mathbf{k},t) = h_{ij}^{*}(-\mathbf{k},t)$, which implies $h_{ij}^{\rightarrow}(\mathbf{k}) = h_{ij}^{\leftarrow *}(-\mathbf{k})$. We then assume that the amplitudes $h_{ij}^{\rightarrow}(\mathbf{k})$ and $h_{ij}^{\leftarrow}(\mathbf{k})$ are Gaussian stochastic processes with zero mean and a given power spectrum and that the integral in eq. 1 exists[22].

As each plane wave is transverse and traceless, in the coordinate system (x,y,z) where $\mathbf{k} = k\left(\sin\theta\cos\phi\, u_x + \sin\theta\sin\phi\, u_y + \cos\theta\, u_z\right)$, we can write

$$h_{ij}^{\rightarrow}(\mathbf{k}) = h_{ij}^{\rightarrow +}(\mathbf{k}) + h_{ij}^{\rightarrow \times}(\mathbf{k}) =$$
$$= A_{ik}(\theta,\phi)\left(h^{\rightarrow +}(\mathbf{k})e_{km}^{+} + h^{\rightarrow \times}(\mathbf{k})e_{km}^{\times}\right)A_{mj}^{-1}(\theta,\phi) \quad [A2]$$

where $e_{ij}^{+} = \begin{pmatrix} 1 & 0 & 0 \\ 0 & -1 & 0 \\ 0 & 0 & 0 \end{pmatrix}$ and $e_{ij}^{\times} = \begin{pmatrix} 0 & 1 & 0 \\ 1 & 0 & 0 \\ 0 & 0 & 0 \end{pmatrix}$ are the two independent polarisation tensors in the TT frame and $A_{ij}(\theta,\phi)$ is the Euler matrix that rotates the axes, from the TT system, where $\mathbf{k} = (0,0,k)$, to the (x,y,z) frame. It follows that

$$h_{ij}^{\rightarrow}(\mathbf{k}) =$$
$$= h^{\rightarrow +}(\mathbf{k})\begin{pmatrix} \cos^2\theta\cos^2\phi - \sin^2\phi & (\cos^2\theta+1)\sin\phi\cos\phi & -\sin\theta\cos\theta\cos\phi \\ (\cos^2\theta+1)\sin\phi\cos\phi & \cos^2\theta\sin^2\phi - \cos^2\phi & -\sin\theta\cos\theta\sin\phi \\ -\sin\theta\cos\theta\cos\phi & -\sin\theta\cos\theta\sin\phi & \sin^2\theta \end{pmatrix} +$$

$$+ h^{\rightarrow \times}(\boldsymbol{k}) \begin{pmatrix} -\cos\theta\sin 2\phi & \cos\theta\cos 2\phi & \sin\theta\sin\phi \\ \cos\theta\cos 2\phi & \cos\theta\sin 2\phi & -\sin\theta\cos\phi \\ \sin\theta\sin\phi & -\sin\theta\cos\phi & 0 \end{pmatrix}, \qquad [A3]$$

i.e., $h_{ij}^{\rightarrow}(\boldsymbol{k}) = h^{\rightarrow +}(\boldsymbol{k})\psi_{ij}^{+}(\theta,\phi) + h^{\rightarrow \times}(\boldsymbol{k})\psi_{ij}^{\times}(\theta,\phi)$ with obvious definitions of the $\psi$'s.

In order to satisfy the reality condition for the transform in eq. 1 we need:
$h_{ij}^{\leftarrow}(\boldsymbol{k}) = h^{\rightarrow +*}(-\boldsymbol{k})\psi_{ij}^{+*}(\pi-\theta,\phi+\pi) + h^{\rightarrow \times *}(-\boldsymbol{k})\psi_{ij}^{\times *}(\pi-\theta,\phi+\pi)$, or, as can be easily checked from eq. A3 by substituting $\theta \rightarrow \pi - \theta$ and $\phi \rightarrow \pi - \phi$,
$h_{ij}^{\leftarrow}(\boldsymbol{k}) = h^{\rightarrow +*}(-\boldsymbol{k})\psi_{ij}^{+}(\theta,\phi) - h^{\rightarrow \times *}(-\boldsymbol{k})\psi_{ij}^{\times}(\theta,\phi)$, so that only two complex amplitudes are really independent and we call them simply $h^{+}(\boldsymbol{k})$ and $h^{\times}(\boldsymbol{k})$.

Assume now that the processes $h^{+}(\boldsymbol{k})$ and $h^{\times}(\boldsymbol{k})$ are uncorrelated, stationary, and isotropic:

$$\langle h^{+}(\boldsymbol{k})h^{\times *}(\boldsymbol{k}')\rangle = \langle h^{+}(\boldsymbol{k})h^{\times}(\boldsymbol{k}')\rangle = 0,$$
$$\langle h^{+}(\boldsymbol{k})h^{+*}(\boldsymbol{k}')\rangle = \langle h^{\times}(\boldsymbol{k})h^{\times *}(\boldsymbol{k}')\rangle = (2\pi)^3 S(k)\delta(k-k')\delta^2(\boldsymbol{u}_k,\boldsymbol{u}_{k'})$$

[A4]

where $S(k) \equiv S(|\boldsymbol{k}|)$, $\boldsymbol{u}_k$ and $\boldsymbol{u}_{k'}$ are the unit vectors parallel to $\boldsymbol{k}$ and $\boldsymbol{k}'$ and $\delta^2(\boldsymbol{u}_k,\boldsymbol{u}_{k'})$ is the Dirac $\delta$ function on the unit sphere. This implies that the two-point, two-time correlation function of the GW stochastic background $\langle h_{ij}(\boldsymbol{r},t)h_{lm}(\boldsymbol{r}',t')\rangle$ is just a function of the modulus of the distance between the two points and of the modulus of the time difference

$$\langle h_{ij}(\boldsymbol{r},t)h_{lm}(\boldsymbol{r}',t')\rangle = \frac{2}{(2\pi)^3} \times$$

$$\times \iiint \cos[kc(t-t')]\left\{\left[\psi_{ij}^{+}(\boldsymbol{k})\psi_{lm}^{+}(\boldsymbol{k}) + \psi_{ij}^{\times}(\boldsymbol{k})\psi_{lm}^{\times}(\boldsymbol{k})\right]S(k)e^{i\boldsymbol{k}(\boldsymbol{r}-\boldsymbol{r}')}\right\}d^3k =$$

[A5]

$$= \frac{1}{(2\pi)^3} \cdot \int_{-\infty}^{\infty} dk k^2 S(k) e^{-ikc(t-t')} \times \int_0^{2\pi} d\phi \int_0^{\pi} \sin\theta d\theta \times$$

$$\times \left[ \psi_{ij}^+(\theta,\phi)\psi_{lm}^+(\theta,\phi) + \psi_{ij}^\times(\theta,\phi)\psi_{lm}^\times(\theta,\phi) \right] e^{ikR[\sin\theta\sin\theta'\cos(\phi-\phi')+\cos\theta\cos\theta']}$$

where we have assumed $\bm{r} - \bm{r}' \equiv R\left(\sin\theta'\cos\phi'\,\bm{u}_x + \sin\theta'\sin\phi'\,\bm{u}_y + \cos\theta'\,\bm{u}_z\right)$.

By taking $\phi'=0$, $\theta'=0$ and $S(\omega) = \frac{8}{15\pi}\frac{k^2}{c}S(k)$, the angular part of the integral in eq. A5 can be performed explicitly and eq. 2 is obtained with the following values for the matrices $\bm{T^0}$, $\bm{T^1}$ and $\bm{T^2}$:

$$\bm{T}^0 = \begin{pmatrix} e_{xz}^+ & \frac{1}{2}e_{xy}^\times & 0 \\ \frac{1}{2}e_{xy}^\times & e_{xz}^+ - e_{xy}^+ & 0 \\ 0 & 0 & e_{xy}^+ - 2e_{xz}^+ \end{pmatrix},$$

$$\bm{T}^1 = \begin{pmatrix} e_{xz}^+ - e_{xy}^+ & -\frac{1}{2}e_{xy}^\times & 0 \\ -\frac{1}{2}e_{xy}^\times & e_{xz}^+ & 0 \\ 0 & 0 & e_{xy}^+ - 2e_{xz}^+ \end{pmatrix}, \qquad [A6]$$

$$\bm{T}^2 = \begin{pmatrix} 0 & 0 & e_{xz}^\times \\ 0 & 0 & e_{yz}^\times \\ e_{xz}^\times & e_{yz}^\times & 0 \end{pmatrix},$$

where $\bm{e}_{xy}^+ = \begin{pmatrix} 1 & 0 & 0 \\ 0 & -1 & 0 \\ 0 & 0 & 0 \end{pmatrix}$, $\bm{e}_{xz}^+ = \begin{pmatrix} 1 & 0 & 0 \\ 0 & 0 & 0 \\ 0 & 0 & -1 \end{pmatrix}$, $\bm{e}_{xy}^\times = \begin{pmatrix} 0 & 1 & 0 \\ 1 & 0 & 0 \\ 0 & 0 & 0 \end{pmatrix}$,

$\bm{e}_{xz}^\times = \begin{pmatrix} 0 & 0 & 1 \\ 0 & 0 & 0 \\ 1 & 0 & 0 \end{pmatrix}$, $\bm{e}_{yz}^\times = \begin{pmatrix} 0 & 0 & 0 \\ 0 & 0 & 1 \\ 0 & 1 & 0 \end{pmatrix}$ and $\bm{0}$ is the null 3×3 matrix.

## Appendix B.

By substituting eq. 16 in eq. 17 we immediately get

$$\sum_{a=1}^{N} \int_{-T}^{T} dt \int_{-T}^{T} dt' g_{aa}(t,t') R_n^a(t-t') +$$
$$+ \sum_{a,b=1}^{N} \int_{-T}^{T} dt \int_{-T}^{T} dt' g_{ab}(t,t') R_s^{ab}(t-t') = 1 \quad [B1]$$

from which eqs. 19b and 20 readily follow.

The $\left\langle \left(\hat{A}^2\right)^2 \right\rangle$ term which enters the estimation of $\sigma_{\hat{A}^2}^2$ can be calculated by using 19b as

$$\left\langle \left(\hat{A}^2\right)^2 \right\rangle = \sum_{a \neq b, c \neq d} \int_{-T}^{T} dt \int_{-T}^{T} dt' \int_{-T}^{T} dt'' \int_{-T}^{T} dt''' g_{ab}(t,t') g_{cd}(t'',t''') \times$$
$$\times \left\langle x^a(t) x^b(t') x^c(t'') x^d(t''') \right\rangle \quad [B2]$$

Then using eq. 13 and the known rule for zero mean Gaussian random variables $\langle xyzw \rangle = \langle xy \rangle \langle zw \rangle + \langle xz \rangle \langle yw \rangle + \langle xw \rangle \langle yz \rangle$ we get

$$\left\langle \left(\hat{A}^2\right)^2 \right\rangle - \left\langle \hat{A}^2 \right\rangle^2 = \sum_{a \neq b, c \neq d} \int_{-T}^{T} dt \int_{-T}^{T} dt' \int_{-T}^{T} dt'' \int_{-T}^{T} dt''' g_{ab}(t,t') g_{cd}(t'',t''') \cdot$$
$$\cdot \Big\{ R_n^a(t''-t) R_n^b(t'''-t') \delta_{ac} \delta_{bd} + R_n^a(t'''-t) R_n^b(t''-t') \delta_{ad} \delta_{bc} +$$
$$+ A^2 \Big[ R_s^{ac}(t''-t) R_n^b(t'''-t') \delta_{bd} + R_s^{ad}(t'''-t) R_n^b(t''-t') \delta_{bc} + \quad [B3]$$
$$+ R_s^{bd}(t'''-t') R_n^a(t''-t) \delta_{ac} + R_s^{bc}(t''-t') R_n^a(t'''-t) \delta_{ad} \Big] +$$
$$+ A^4 \Big[ R_s^{ac}(t''-t) R_s^{bd}(t'''-t') + R_s^{ad}(t'''-t) R_s^{bc}(t''-t') \Big] \Big\}$$

Assume now that the correlation signal $As^a(t)$ is negligible in comparison to the intrinsic noise $\eta^a(t)$, as is the case in the real physical situation. Hence, in eq. B3 we can neglect the terms containing $A^2$ and $A^4$, getting

$$\sigma^2_{\hat{A}^2} = \left\langle \left(\hat{A}^2\right)^2 \right\rangle - \left\langle \hat{A}^2 \right\rangle^2 \approx 2 \sum_{a \neq b} \int_{-T}^{T} dt \int_{-T}^{T} dt' \int_{-T}^{T} dt'' \int_{-T}^{T} dt''' \times$$

$$\times g_{ab}(t,t') g_{ab}(t'',t''') R_n^a(t''-t) R_n^b(t'''-t')$$
[B4]

The problem reduces to a constrained variational problem where, with the help of the standard lagrangian multiplier technique, we minimise $\sigma^2_{\hat{A}^2}$ under the constraint of eq. B1. The functional

$$\Lambda(g_{ab},\lambda) \equiv \sum_{a \neq b} \left\{ 2 \int_{-T}^{T} dt \int_{-T}^{T} dt' \int_{-T}^{T} dt'' \int_{-T}^{T} dt''' \times \right.$$

$$\times \left[ g_{ab}(t,t') g_{ab}(t'',t''') R_n^a(t''-t) R_n^b(t'''-t') \right] +$$
[B5]

$$\left. + \lambda \int_{-T}^{T} dt \int_{-T}^{T} dt' g_{ab}(t,t') R_s^{ab}(t'-t) \right\}$$

reaches its minimum when

$$\frac{\delta \Lambda(g_{ab},\lambda)}{\delta g_{ab}} = 0 \qquad 1 \leq a, b \leq N \text{ and } a \neq b,$$
[B6]

i.e., when eq. 19a is obeyed.

# References

[1] M. Gasperini, M. Giovannini and G. Veneziano, Phys. Rev. Lett. **75**, 3796 (1995)

[2] P. F. Michelson, Mon. Not. R. astr. Soc. **227**, 933 (1987)

[3] E.E. Flanagan, Phys Rev **D48**, 2389 (1993)

[4] B. Allen in proc. of the Les Houches Summer School on "Astrophysical Sources of Gravitational Radiation". J. A. Marck and J. P. Lasota eds. (World Scientific Singapore, in press)

[5] NAUTILUS is at the Laboratori Nazionali di Frascati (Rome); see: E.Coccia et al. in "Gravitational Wave Experiments". E. Coccia, G. Pizzella and F. Ronga eds. (World Scientific Singapore 1995) p. 161. AURIGA is at the Laboratori Nazionali di Legnaro (Padua) of the Istituto Nazionale di Fisica Nucleare; see: M. Cerdonio et al. *ibidem* p. 176.

[6] VIRGO is a large interferometric antenna under construction near Pisa, Italy. See: A. Giazotto et al. in "Gravitational Wave Experiments". E. Coccia, G. Pizzella and F. Ronga eds. (World Scientific Singapore 1995) p.86.

[7] See for instance ref. 4 and references therein.

[8] Y Gürsel and M. Tinto, Phys. Rev. **D40**, 3884 (1989)

[9] C. Z. Zhou and P. Michelson, Phys. Rev. **D51**, 2517 (1995)

[10] S. M. Merkowitz and W. J. Johnson, Phys. Rev. **D51**, 2546 (1995)

[11] R. Giffard, Phys. Rev. **D14**, 2478 (1976)

[12] ALLEGRO at the University of Louisiana, Baton Rouge, EXPLORER at Cern, Geneva and NIOBE at University of Western Australia, Perth.

[13] A Giazotto, Phys. Rep. **182**, 365 (1989) and A Giazotto, Proc. of the I VIRGO meeting, Pisa 1996 ,F. Fidecaro ed, (World Scientific, Singapore, in press)

[14] E. Coccia, A. Lobo and J.A. Ortega, Phys. Rev. **D52**, 3735 (1995)

[15] INFN Laboratori Nazionali del Gran Sasso.

[16] See for instance: P. Astone, J.A. Lobo and B. Shultz, Class. Quantum Grav. **11**, 2093 (1994).

[17] C.L. Bennet et al Astrophys. J. **436**, 423 (1994)

[18] S. E. Thorsett and R. J. Dewey, Phys. Rev. **D53**, 3468 (1996).

[19] S.P. Boughn, S. J. VanHook and C.M. O'Neill, Astrophys. J. **354**, 406 (1990)

[20] J. Hough, J. R. Pugh, R. Bland and R. W. Drever, Nature **254**, 498 (1975).

[21] P. Astone et al., (R.O.G. collaboration) Submitted to Phys. Lett. (1996)

[22] This implies that the integral has to be cut off at a maximum value $k_{max}$.

**Caption to Figures**

Fig 1 Correlation function $\Theta(\theta,x)$ for two parallel bar detectors as a function of the reduced distance $x=\omega R/c$, with R the detector distance. The different curves are parametrized by the value of the angle $\theta$ between the detector axis and the line joining the detector sites. The flattest curve corresponds to $\theta=0$, the curve of maximum oscillation is for $\theta=\pi/2$. The dotted line corresponds to $\theta=1$ rad, which is the orientation of the AURIGA -NAUTILUS pair.

Fig 2 Correlation function $\Xi(m,x)$ for the $-2\leq m \leq 2$ modes of two spheres as a function of $x=\omega R/c$, with R the detectors distance. The different curves refer to the different modes according to the legend in the insert. The modes relate to a reference frame where the z axis is along the line joining the detectors

Fig. 3 The overall correlation function $\sqrt{\sum_{m=-2}^{2}\Xi^2(m,x)}$ for two spheres as a function of the reduced distance $x = \omega R/c$.

Fig. 4. The total correlation function $A(x)=\sqrt{\sum_{m=1}^{5}\Phi^2(m,x)}$ for an interferometer and a sphere as a function of the reduced distance $x = \omega R/c$. Here the function $\Phi(m,x)$ is the correlation function between the interferometer and the m$^{th}$ mode of the sphere.

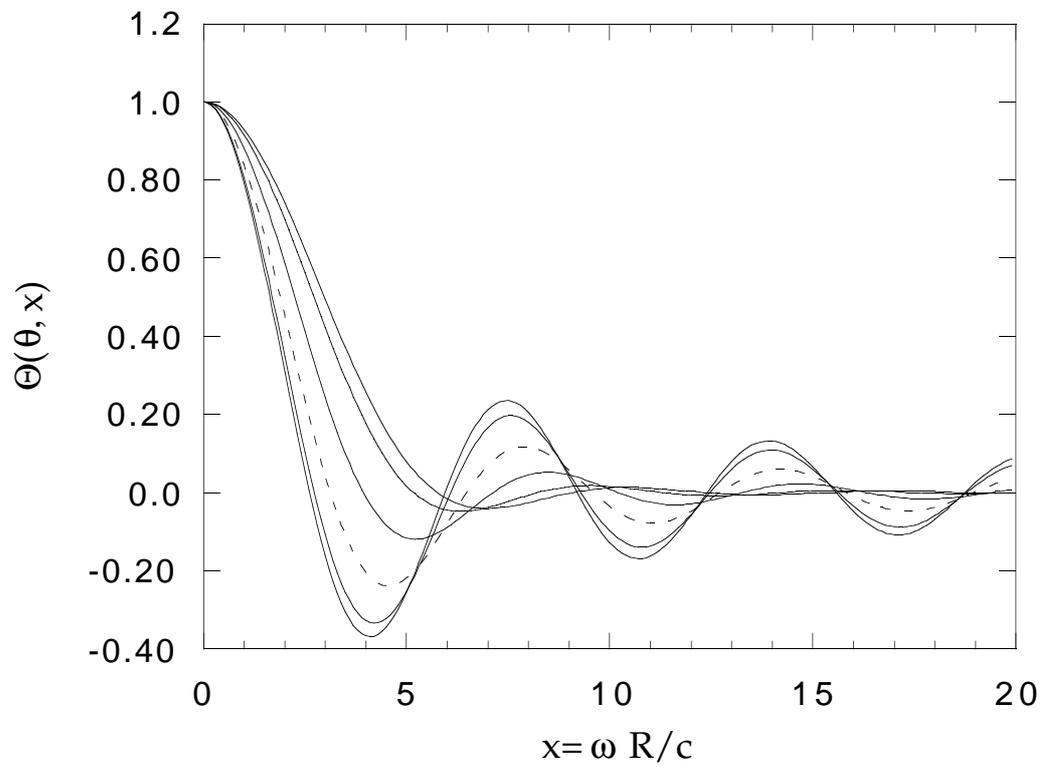

Fig. 1

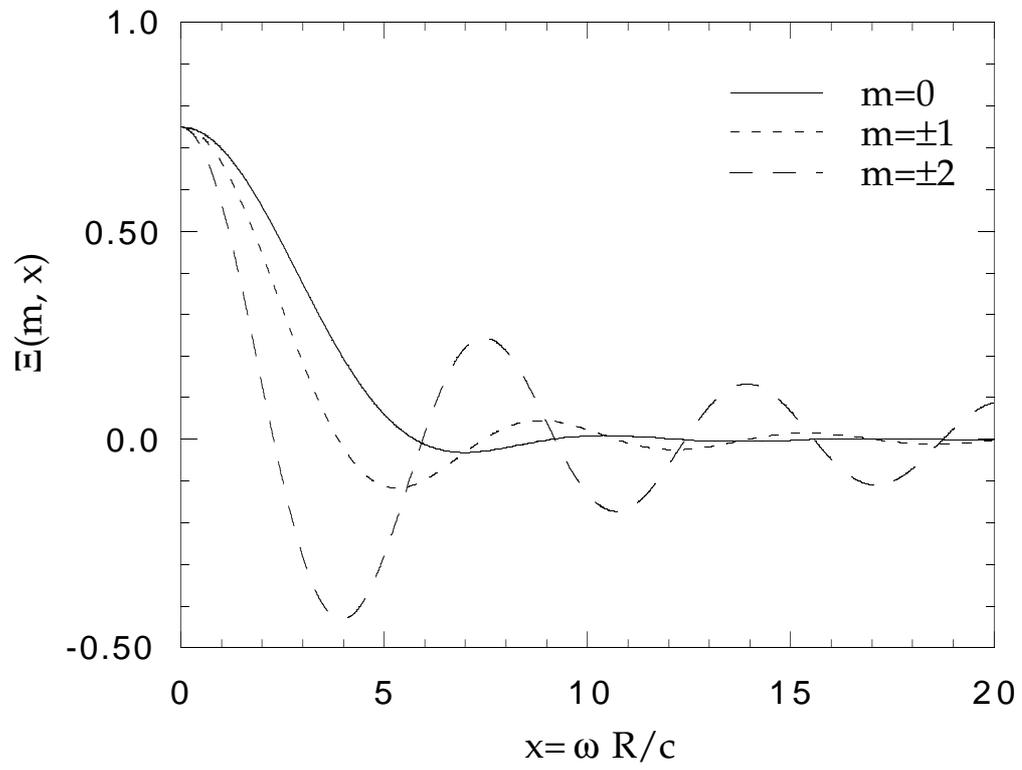

Fig. 2

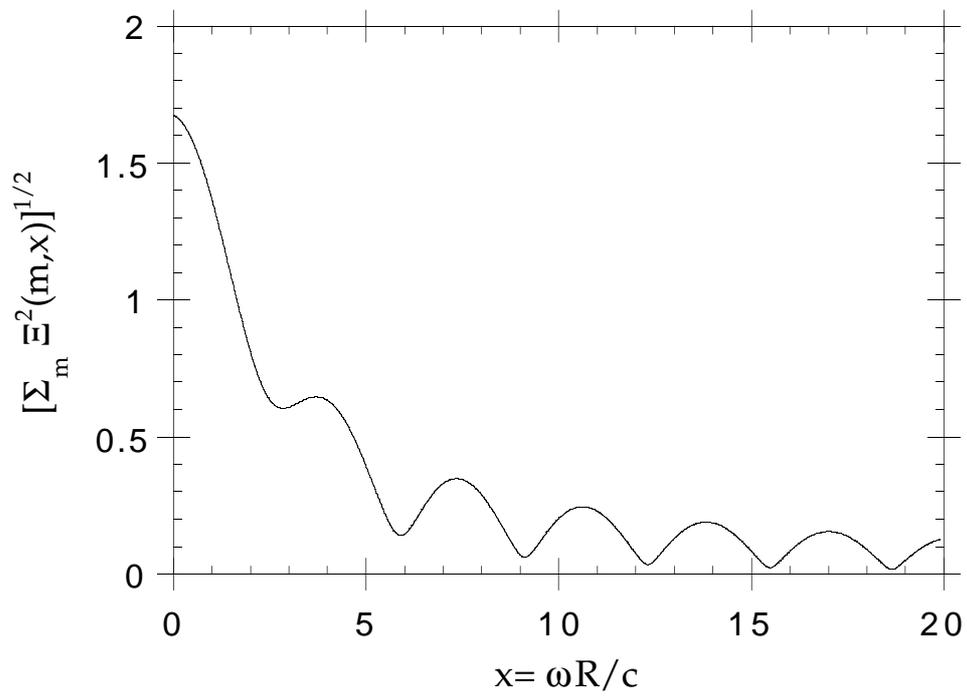

Fig. 3

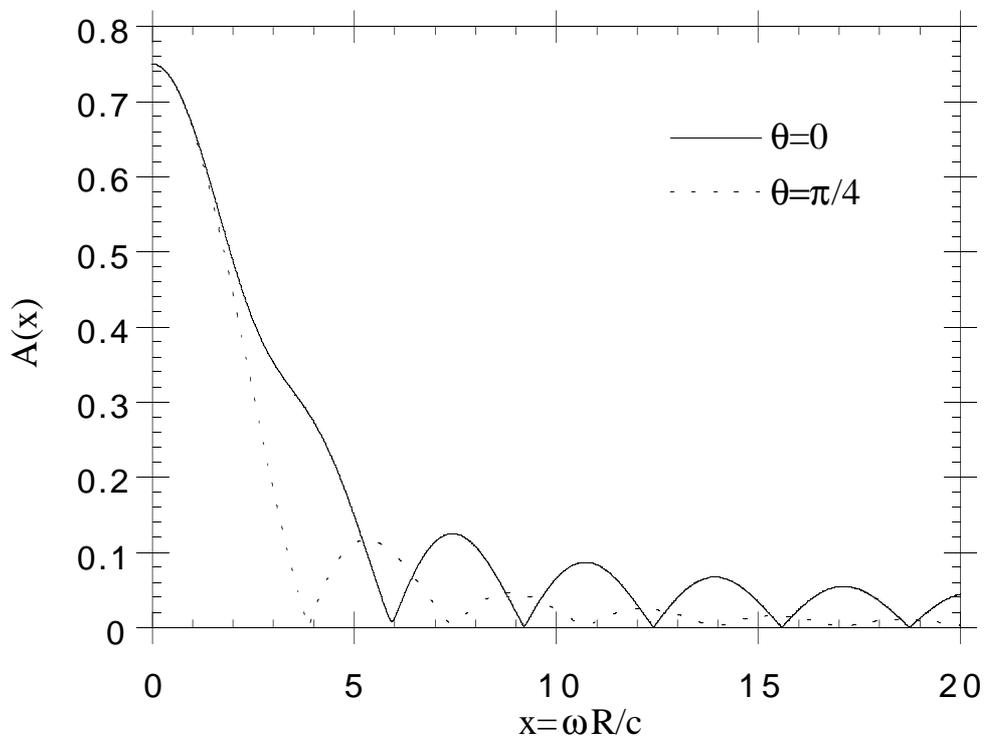

Fig. 4